\documentclass[aps,prl,twocolumn,superscriptaddress]{revtex4}

\usepackage{graphicx}
\usepackage{dcolumn}
\usepackage{bm}
\makeatletter
\def\captionof#1#2{{\def\@captype{#1}#2}}
\makeatother

\begin{document}

\title{Driven quantum coarsening}

\author{Camille Aron}
\affiliation{Universit{\'e} Pierre et Marie Curie - Paris VI,
LPTHE UMR 7589, 4 Place Jussieu, 75252 Paris Cedex 05, France}
\author{Giulio Biroli}
\affiliation{
Institut de Physique Th\'eorique,
CEA, IPhT, F-91191 Gif-sur-Yvette, France
CNRS, URA 2306}
\author{Leticia F. Cugliandolo}
\affiliation{Universit{\'e} Pierre et Marie Curie - Paris VI,
LPTHE UMR 7589, 4 Place Jussieu, 75252 Paris Cedex 05, France}

\begin{abstract}
  We study the driven dynamics of quantum coarsening. We analyze
  models of $M$-component rotors coupled to two electronic reservoirs
  at different chemical potentials that generate a current threading
  through the system. In the large $M$ limit we derive the dynamical
  phase diagram as a function of temperature, strength of quantum
  fluctuations, voltage and coupling to the leads. We show that the
  slow relaxation in the ordering phase is universal. On large time
  and length scales the dynamics are analogous to stochastic
  classical ones, even for the quantum system driven out of equilibrium
  at zero temperature. We argue that our results apply to generic
  driven quantum coarsening.
\end{abstract}

\pacs{Valid PACS appear here}
\maketitle

Phase transitions are central to condensed matter and
statistical physics.  Initially, emphasis was put on classical and
quantum {\it equilibrium} phase transitions.
Later, attention moved to {\it non-equilibrium} phase transitions in
which quantum fluctuations can be neglected. These are realized when a
system is forced in a non equilibrium steady state (by a shear rate,
an external current flowing through it,
etc.)~\cite{OnukiKawasaki,DombGreen} or when it just fails to relax
(e.g. after a quench) and displays aging
phenomena~\cite{Struick,LeticiaLesHouches}.  The study of steady
states in small quantum systems driven out of
equilibrium~\cite{MitraMillisReichman} has been recently boosted by
their relevance for nano-devices.  In contrast, the effect of a drive
on a {\it macroscopic} system close to a quantum phase transition is a
rather unexplored subject. Some works have focused on non-linear
transport properties close to an (equilibrium) quantum phase
transition~\cite{DalidovichPhillips,GreenSondhi,HoganGreen}. Others
have studied how the critical properties are affected by a drive
\cite{MitraKimMillis,Feldman,MitraMillis}.  However, a global
understanding of phase transitions in the parameter space $T$
(temperature), $V$ (drive), $\Gamma$ (strength of quantum
fluctuations), and the r{\^o}le played by the environment, is still
lacking. Furthermore, experiments in $2d$ electronic
systems~\cite{Ovadyahu,Popovic} show interesting features in the {\it
relaxation} toward the quantum non-equilibrium steady state (QNESS)
but these have not been addressed theoretically yet
 (except for \cite{mueller}).

 A number of intriguing questions arise in the context of driven
quantum phase transitions, some of which are: How long does it take to
reach the QNESS after one of the parameters $T,\ V,\ \Gamma$ is
changed?  Do the systems always relax to the QNESS or, as for classical
systems, do quenches deep in the $T,\ V, \ \Gamma$ phase diagram lead
to aging phenomena and glassy dynamics? What are the properties
of the latter `doubly non-equilibrium' dynamics? Are quantum quenches,
obtained by changing $V$ and $\Gamma$ at $T=0$, different from their
classical counterpart?\\
\begin{figure}
\begin{minipage}[]{0.45\linewidth}
\includegraphics[width=3.5cm]{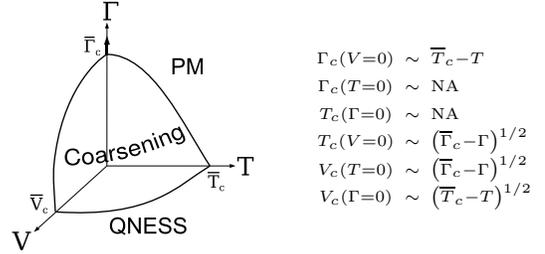}
\end{minipage}
\begin{minipage}[]{0.45\linewidth}
\begin{small}
\begin{tabular}{ccl}
$\scriptstyle \Gamma_c(V=0)$ & $\scriptstyle\sim$ & $\scriptstyle\overline{T}_c - T$\\
$\scriptstyle \Gamma_c(T=0)$ & $\scriptstyle\sim$ & $\scriptstyle{\rm NA}$\\
$\scriptstyle T_c(\Gamma=0)$ & $\scriptstyle\sim$ & $\scriptstyle{\rm NA}$\\
$\scriptstyle T_c(V=0)$ & $\scriptstyle\sim$ & $\scriptstyle\left( \overline{\Gamma}_c - \Gamma \right)^{1/2}$\\
$\scriptstyle V_c(T=0)$ & $\scriptstyle\sim$ &$\scriptstyle\left( \overline{\Gamma}_c - \Gamma \right)^{1/2}$\\
$\scriptstyle V_c(\Gamma=0)$ & $\scriptstyle\sim$ & $\scriptstyle\left( \overline{T}_c - T \right)^{1/2}$
\end{tabular}
\end{small}
\end{minipage}\hfill

\caption{\label{fig:epsart} Non-equilibrium phase diagram for
reservoirs with a much larger bandwidth than all other energy scales
($\hbar\omega_F\gg J$). Close to the critical point $\overline
V_c=V_c(T=\Gamma=0)$ the critical lines are non-analytical (${\rm NA}$). Close to
$\overline T_c$ and $\overline\Gamma_c$ they are power laws with
exponents given next to the figure. The arrow above $\Gamma_c$ 
indicates that the critical surface is pulled up by increasing the 
coupling to the leads.}
\end{figure}

 The aim of this work is to answer these questions for a
class of analytically tractable models, systems of $M-$component
quantum rotors that encompass an infinite range spin-glass and its
3d pure counterpart modeling coarsening phenomena.
Models of quantum rotors are non-trivial but still relatively simple
and provide a coarse-grained description of physical systems such as
Bose-Hubbard models and double layer
antiferromagnets~\cite{SachdevBook}. The out of equilibrium 
drive is provided by 
two external electron reservoirs that induce a
current flowing through the system. 
In the simplest setting~\cite{MitraKimMillis} each rotor is coupled to
two independent reservoirs. Using the Schwinger-Keldysh formalism we analyze
the out of equilibrium dynamics in the large $M$ limit. We find a
phase transition, see Fig.~1,
between a QNESS ($V\neq0$) and an ordered phase, we study its 
critical properties
and we discuss the effect of the environment.

 The model we focus on is an infinite-range quantum
disordered system made of $N$ $M$-component rotors interacting via
random Gaussian distributed couplings, $J_{ij}$, with zero mean and
variance $J^2/N$.  Its Hamiltonian is
\begin{equation} 
H_S=\frac{\Gamma}{2\hbar^2 M}\sum_{i=1}^N\mathbf{L}_i^2 - \sum_{i<j}J_{ij}\mathbf{n}_i
\mathbf{n}_j \; ,  \;\;\; \mathbf{n}_i^2=M \;\; \forall i\; .
\label{eq:model} 
\end{equation}
$n_{i}^\mu$ are the $M$ components of the $i$-th rotor and
$L_i^{\mu\nu}=n_i^\mu p_i^\nu-n_i^\nu p_i^\mu$, with
$p_i^\mu=-i\hbar\partial/\partial n_i^\mu$, are the $M(M-1)$ components of
the $i$-th generalized angular momentum operator with $\mathbf{L}_i^2
= \sum_{\mu<\nu} (L_i^{\mu\nu})^2$~\cite{SachdevBook,SachdevYe}.
$\Gamma$ controls the
strength of quantum fluctuations; as $\Gamma\to 0$ the model
approaches the classical $M$-component Heisenberg fully-connected
spin-glass.  In the large $M$ limit it is equivalent to the quantum
fully-connected $p=2$ spherical spin-glass~\cite{Cude,Rokni}.  A
mapping to ferromagnetic coarsening in the $3d$ O(${\cal
N}$) model can be established in the classical and large ${\cal N}$
limits \cite{LeticiaLesHouches}.  As we shall show, this mapping 
holds for the quantum model as well. Thus, it allows us to extend our results
also to the 3d ferromagnetic model in the large $M$ limit.

The system is coupled to two independent and non-interacting `left'
($L$) and `right' ($R$) electronic reservoirs in equilibrium at
different chemical potentials $\mu_L=\mu$ and $\mu_R=\mu+eV$ and the
same temperature $T$.  $R$ and $L$ reservoirs act as source and drain,
respectively. The details of their Hamiltonian are not important
since in the small spin-bath coupling we concentrate on 
only the electronic Green
functions matter. We focus on free electrons with the same symmetric
density of states, $\rho_L(\epsilon)=\rho_R(\epsilon)=\rho(\epsilon)$, centered at $\epsilon=0$, and
with typical variation scale around $\hbar\omega_F$ .

Each rotor is coupled non-linearly to its (double) electron bath. 
For example, for $M=3$ rotors we take
$
H_{SB}=  \frac{1}{{\cal M}} \sum_{i\alpha l l'\gamma\gamma'}
\hbar \omega_{\gamma\gamma'} n_i^\alpha \; 
[c^{\dagger}_{liL\gamma}\frac{\sigma_{ll'}^\alpha}{2}c_{l'iR\gamma}
+
L \leftrightarrow R
]
$
where $c^{\dagger}_{liR(L)\gamma},\ c_{liR(L)\gamma}$ are the fermionic
operators of the $R(L)$ reservoirs, $\hbar \omega_{\gamma\gamma'}$ is the
electron-bath coupling, chosen to be constant: $\hbar \omega_{\gamma\gamma'}=\hbar\omega_c$.
$\mathbf{\sigma}^{\alpha}$ are the Pauli matrices ($\alpha=1,...,M$). $\gamma=1,...,\cal
M$ is the fermion label inside the reservoirs.

System and reservoirs are uncoupled at time $t<0$ and 
evolve with $H=H_S+H_B+H_{SB}$ at $t>0$. The density matrix at $t=0$,
$\varrho=\varrho_S \otimes \varrho_L \otimes \varrho_R$,
provides the initial condition.  $\varrho_S$,
$\varrho_L$ and $\varrho_R$ correspond to equilibrium of the system at
temperature $T_0\gg 1$, and the $L$ and $R$ reservoirs at temperature 
$T$ and chemical potentials $\mu$ and $\mu+eV$, respectively.  For simplicity,
$\varrho_S$ is taken to be the identity; this choice is
equivalent to any other one uncorrelated with disorder~\cite{Culo}.

We analyze the $t>0$ dynamics by using the Schwinger-Keldysh formalism
yielding a functional-integral representation of the Heisenberg
evolution \cite{SchwingerKeldysh,Culo,preparation}.  Each field carries a $\pm$
index associated to the forward and backward evolution. 
The action corresponding to Eq.~(\ref{eq:model}) is
\begin{displaymath}
S_S=\sum_{a=\pm}
a
\int {\rm d}t 
\left[ 
\frac{\hbar^2}{2\Gamma}
\sum_i
(\mathbf{\dot n}_{ia})^2
+
\sum_{i<j}J_{ij}\mathbf{n}_{ia}
\mathbf{n}_{ja} 
\right]
\; .
\end{displaymath}
The path-integral runs over paths such that $\mathbf{n}_{ia}^2(t)=M, \
\forall iat$.
One may lift this constraint by using 
the integral representation of the Dirac delta.
This amounts to introducing auxiliary imaginary fields 
$\lambda_{ia}(t)$ and adding
$S_\lambda=\sum_{a=\pm}\frac{a}{2}\int {\rm d}t \sum_{i}
\lambda_{ia}(t)(\mathbf{n}_{ia}^2(t)-M)$ to the action.\\
After expanding the system-leads interaction up to second order in
$g\equiv\omega_c/\omega_F$, integrating out the fermionic fields, and
taking the large $M$ limit we obtain a
(Feynman-Vernon-like) action for the rotors. The detailed
computation~\cite{preparation} confirms that several system-reservoir
coupling that preserve the $O(M)$ symmetry and the addition of
different $LL$ and $RR$ couplings do not modify our results
qualitatively. In short we obtain
\begin{eqnarray*}
&& S_{SB}=- \frac 1 2 \sum_{ab=\pm}\int {\rm d}t {\rm d}t' 
\ \Sigma_{ab}^B(t,t') \ \sum_{i} \mathbf{n}_{ia}(t)\mathbf{n}_{ib}(t')
\; , 
\nonumber\\
&& \Sigma_{ab}^B(t,t')=-iab \hbar \omega_c^2
\left[G^R_{ab}(t,t')G^L_{ba}(t',t)+
 L\leftrightarrow R
\right]
\; . 
\end{eqnarray*}
The electronic Green functions are 
$G_{ab}(t,t') \equiv -i \langle {\cal T} \psi_a(t) \psi^\dag_b(t')\rangle$
with $\psi_a(t), \psi^\dag_a(t)$ the fermionic fields
and ${\cal T}$ the time-ordering operator on the closed contour. 
It is convenient to change basis and use retarded,
$G_R^B=(G_{+-}^B-G_{++}^B)/\hbar$, advanced, $G_A^B=(G_{-+}^B-G_{++}^B)/\hbar$, and
Keldysh, $G_K^B=i(G_{++}^B+G_{--}^B)/2$, Green functions. The $\Sigma$'s
transform in a similar way.  For identical reservoirs at temperature
$T$ and chemical potential $\mu$ ($V=0$), the self-energy
components verify the usual fluctuation dissipation relation of a
standard {\it bosonic} bath $\Sigma_K^B(\omega)= \hbar
\coth(\beta\hbar\omega/2) \Im \Sigma_R^B(\omega)$, with 
$\beta=1/T$ and $k_B=1$.

Collecting all contributions the total action,
$S=S_S+S_\lambda+S_{SB}$, is $O(MN)$. Given that the zero-source
generating functional equals one, one can simply compute its average
over quenched randomness~\cite{Culo} and use a saddle-point evaluation
of the resulting path-integral that is exact in the large $M$ and $N$
limits. The value of $\lambda_{ia}(t)$ at the saddle point 
is a spatially homogenous function $\lambda(t)$. Its
time-dependence is determined by the condition $\langle
\mathbf{n}_{ia}^2 (t)\rangle =M$ with the average taken over $S$ \cite{preparation}.

The macroscopic dynamic order parameters are the
symmetric two-time correlation and instantaneous linear response that
in the operator formalism are defined as $MC(t,t_w) \equiv \langle
\{ {\bf n}_i(t), {\bf n}_i(t_w)\}/2\rangle$ and $MR(t,t_w) \equiv \delta
\langle {\bf n}_i(t) \rangle/\delta {\bf h}_i(t_w)|_{h=0} =-1/\hbar
\langle [{\bf n}_i(t),{\bf n}_i(t_w)]\rangle \theta(t-t_w)$. The field
${\bf h}_i$ couples linearly to the $i$-th rotor and the last identity
is the Kubo formula valid in linear response. The exact Schwinger-Dyson
equations then read:
\begin{eqnarray}
&&{\cal D}(t)
R(t,t_w)
= \delta(t-t_w) + \int_{t_w}^t {\rm d}t'' \ \Sigma_R(t,t'') R(t'',t_w) 
\; , 
\nonumber\\
\nonumber\\
&&
{\cal D}(t)
C(t,t_w)
= \int_{0}^{t} {\rm d}t'' \ \Sigma_R(t,t'') C(t'',t_w) 
\nonumber\\
&& \qquad \qquad \qquad \;\;\;
+ \int_{0}^{t_w} {\rm d}t'' \ \Sigma_K(t,t'') R(t_w,t'')
\; , \label{SD}
\end{eqnarray}
with ${\cal D}(t)= \hbar^2 \Gamma^{-1} \partial_t^2 + \lambda(t)$, the 
retarded and Keldysh self-energies given by $\Sigma_R=\Sigma_R^B+J^2 R$ and
$\Sigma_K=\Sigma_K^B+$$J^2 C$, and 
$\lambda(t) = -\hbar^2 \Gamma^{-1} \partial^2_{t^2} C(t,t_w\to t^-) +$$
\int_{0}^t {\rm d}t'' \; [ \Sigma_R(t,t'') C(t,t'') +$$  \Sigma_K(t,t'')
R(t,t'')]$.

In the QNESS, the dynamics are stationary but they do not satisfy the
fluctuation-dissipation theorem (FDT) when $V\neq 0$. 
The Lagrange multiplier approaches a constant,
$\lambda(t)\to\lambda_\infty$, and the linear response satisfies a closed
equation that once Fourier transformed reads $R(\omega) = -[\hbar^2 \Gamma^{-1}
\omega^2 - \lambda_\infty + \Sigma_R^B(\omega)+J^2R(\omega)]^{-1}$.
The physical solution to this quadratic equation is the one that satisfies 
$R(\omega \to \infty)=0$.
The correlation is given
by $C(\omega)=\Sigma_K(\omega) |R(\omega)|^2$
and the spherical constraint, $C(t,t)=1$, implies
\begin{equation}
\int_0^\infty \frac{{\rm d}\omega}{2\pi} \; 
\frac{\Sigma_K^B(\omega)}{\Im \Sigma_R^B(\omega)} \; \Im R(\omega) = 1/2
\; . 
\label{eq:critical}
\end{equation}
The phase transition occurs when $R(\omega=0)=\int {\rm d}t R(t) $ ceases to 
be real, indicating that the stationary condition necessary to Fourier transform is no longer valid. Concomitantly, the derivatives of $R(\omega)$ in
$\omega=0$ diverge and hence the real-time response function
shows a power law decay. 
This happens when
$\lambda_\infty=\lambda_c=2J+\Sigma_R^B(\omega=0)$.  Inserting
$\lambda_c$ in Eq.~(\ref{eq:critical}), we then obtain the equation for the
critical manifold in the $T,V,\Gamma$ space (for a given $g$ and
$\hbar \omega_F$). We shall derive the critical manifold for different
reservoirs in full detail in~\cite{preparation}; we summarize here 
some of the salient features.

We first consider $g\to 0$ {\it after} the long-time limit such that
the asymptotic regime has been established and we take $\hbar\omega_F$
much larger than any other energy scale.  For $\Gamma= V=0$ we recover
the classical critical temperature, $\overline
T_c=J$~\cite{SachdevYe,Cude}.  At $V=T=0$ we obtain $\overline
\Gamma_c=(3\pi/4)^2 J$, as for the $p=2$ quantum spherical model in
equilibrium~\cite{SachdevYe} and its dynamics coupled to an
equilibrium oscillator bath~\cite{Rokni}.  Finally, the critical
point $\overline V_c$ on the $\Gamma=T=0$ line is determined by
\begin{eqnarray*}
&&\int_{\mu}^{\mu+e\overline V_c} 
{\rm d}\epsilon \; \left[\frac{\rho_L(\epsilon)}{2J} -\dot\rho_L(\epsilon) \right]
\rho_R(\epsilon)= 
\rho_L\left(\mu\right)
\rho_R\left(\mu\right) 
\end{eqnarray*}
that can be solved numerically and also analytically in some
special cases. In the large bandwidth limit 
$\hbar \omega_F \gg J$ and we find $e\overline V_c=2J$ for the 
$\rho_L=\rho_R=\rho$ symmetric case, with $\mu=0$ and 
differentiable at the origin. The form of the critical lines
are shown in Fig.~1.

As for finite $\hbar\omega_F$ we find that $\overline V_c$ varies 
(contrary to $\overline T_c$ and $\overline \Gamma_c$) upon 
decreasing $\hbar\omega_F/J$, the
critical line $V_c(T,\Gamma=0)$ is re-entrant and, for a single 
band, $eV$ is bounded when the $R$ reservoir is filled.

When the coupling to the electronic reservoirs, $g$, is finite the critical
line in the $\Gamma=0$ plane remains unaltered but the critical surface
on the $\Gamma$ direction is pulled `upwards' enlarging the low temperature
phase for increasing values of $g$. This is similar to what was found
for quantum oscillator Ohmic baths and is due to a spin-localization-like 
effect~\cite{effect-bath,Rokni}.  \\
\begin{figure}
\centerline{
\hspace{0.25cm}
\includegraphics[width=3.5cm,angle=-90]{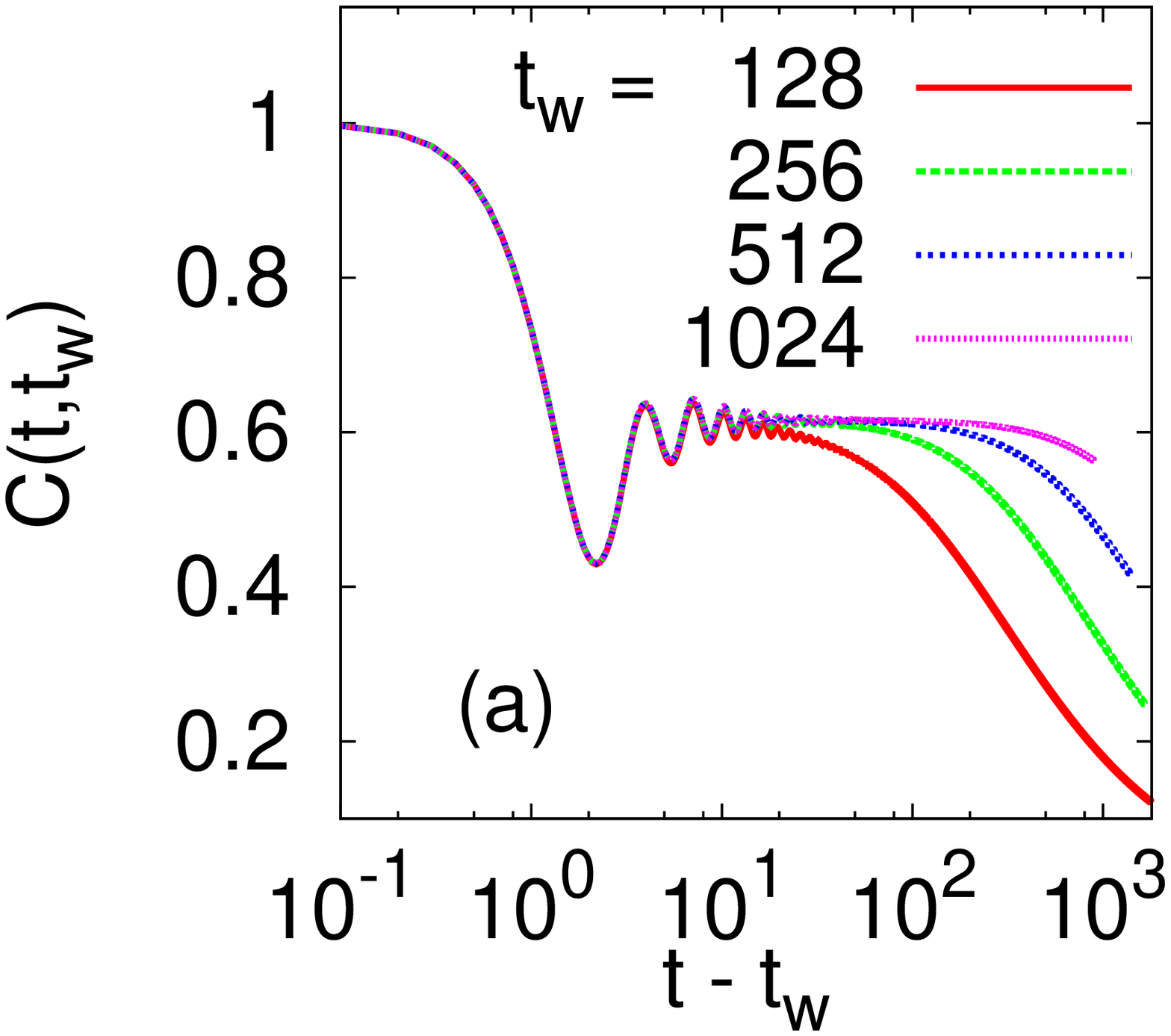}
\hspace{-1.5cm}
\includegraphics[width=3.25cm,angle=-90]{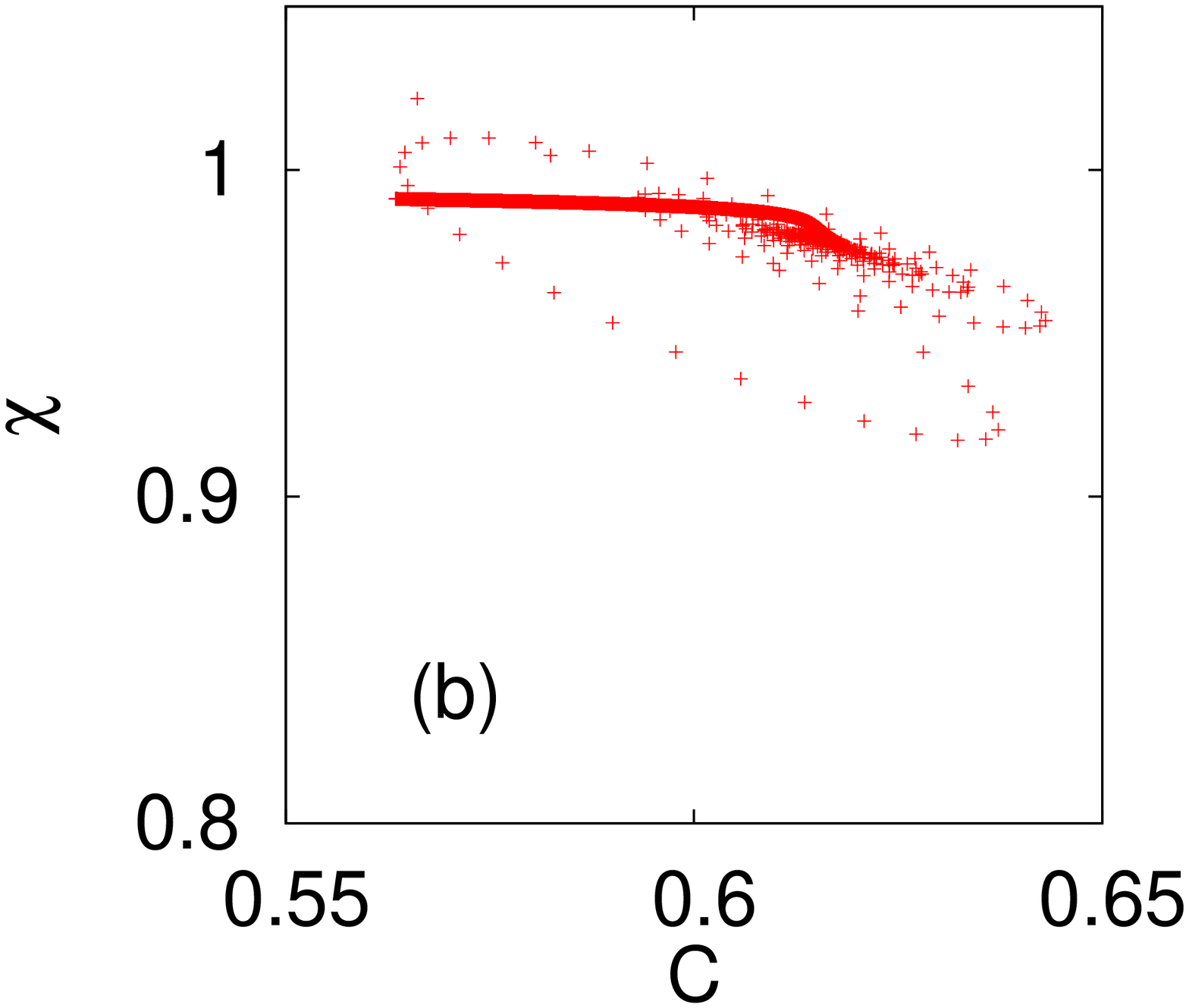}
\hspace{.75cm}
}
\caption{\label{fig:C-chi} Dynamics in the driven coarsening regime:
numerical solution to Eqs.~(\ref{SD}) after a quench to $T=0.2 J,V=0.2
J$, $\Gamma=1$, $g=1$ with $\hbar\omega_F = 10 J$.  (a) The symmetric
correlation $C(t,t_w)$.  (b) The integrated linear response,
$\chi(t,t_w) =\int_{t_w}^t {\rm d}t' R(t,t')$ against $C$, for
$t_w=1024$ and using $t$ as a parameter. The curved part corresponds
to the stationary and oscillatory regime with $(t-t_w)/t_w\to 0$ while
the straight line is for times in the monotonic aging decay of $C$.}
\end{figure}
We now turn to the dynamics.  Our numerical and analytical analysis of
Eqs.~(\ref{SD}) show that after a quench in the low-$T$, weak-$\Gamma$
and weak-$V$ phase, the dynamics do not 
reach a QNESS \cite{preparation}.  There is a separation of two-time
scales typical of aging phenomena \cite{LeticiaLesHouches}. First, a
stationary regime for short time differences $t-t_w$ with respect to
the waiting-time after the quench, $t_w$, in which the symmetric
correlation approaches a plateau asymptotically in the
time-difference. Later, an aging regime in which $C$ depends on the
two times explicitly. This behavior is shown in Fig.~\ref{fig:C-chi}(a).
The plateau value $q_{\rm EA}$, so-called Edwards-Anderson
parameter, measures the fraction of frozen rotor fluctuations on
timescales much smaller than $t_w$.  The stationary decay depends on
all control parameters. $q_{\rm EA}$ approaches one at $T=\Gamma=V=0$ and
zero on the critical manifold as in a second order transition. 
In the aging regime, the
correlation normalized by $q_{\rm EA}$ is identical to the classical
one~\cite{Cude}:
\begin{equation}
C(t,t_w)/q_{\rm EA} \simeq 2\sqrt{2}\ (t_w/t)^{3/4} \ (1+t_w/t)^{-3/2}
\label{eq:C}
\end{equation}
for $0<t/t_w<1$.
We shall prove this result and unveil the connection with coarsening anticipated previously by exploiting the quadradic form of the full action in the ${\bf n}$ fields.
Under the Keldysh rotation $({\bf n}^+,{\bf n}^-) \to (i\hat {\bf
n},{\bf n})$ with $i\hat {\bf n}\equiv i({\bf n}^+-{\bf n}^-)/\hbar$
and ${\bf n}\equiv ({\bf n}^++{\bf n}^-)/2$ the action is identical to
the Martin-Siggia-Rose one for a {\it classical} Langevin process in a
harmonic potential $\sum_{ij}(J_{ij}-\lambda(t)\delta_{ij}){\bf
n}_i\cdot {\bf n}_j$.  The noise statistics is, however, peculiar:
because of the quantum origin of the environment it has memory, 
depends on $\hbar$ and satisfies the quantum FDT in the $V=0$
case. The Langevin equations are rendered independent -- apart from a
residual coupling through the Lagrange multiplier -- by a rotation
onto the basis that diagonalizes the interaction matrix $J_{ij}$:
${\bf n}_\mu=\sum_j v_\mu^j {\bf n}_j$ and $i \hat {\bf n}_\mu=\sum_j
v^j_\mu i \hat {\bf n}_j$ with $v^j_\mu$ the eigenvector associated to
the eigenvalue $J_\mu$.  The analysis then follows the same route as in
\cite{Cude}, see~\cite{preparation}. One finds quite naturally that
the long-time dynamics correspond to a Bose-Einstein-like condensation
process of the $M$ $N$-dimensional `vectors' ${\bf n}_i$ on the
direction of the edge eigenvector. The relaxation is then controlled
by the decay of $\rho(J_\mu)$ close to its edge.  For Gaussian
i.i.d. couplings $\rho(J_\mu) \propto (2J-J_\mu)^{1/2}$. This coincides
with the distribution of the modulus of the Laplacian eigenvalues,
$dk^2 k^{2(d/2-1)}=dk^2 k^{2\times 1/2}$ in $d=3$.  For this reason
all models with a square root singularity of the distribution of
`masses' $J_\mu$, as the ferromagnetic rotor model in $d=3$ and the
completely connected spin glass rotor model, are characterized by the
same long-time dynamics. Now let us show that the {\it aging} dynamics
are indeed equivalent to their classical counterpart.  In the ordered
phase, taking the long $t_w$ {\it and} $t-t_w$ limits with $t/t_w$
fixed (low frequency aging regime) the second-time derivatives in the effective 
Langevin equations can be
neglected. Furthermore, only the low-frequency
($\omega\ll 1/{\beta\hbar}$) behavior of the kernels plays a r\^ole
in this regime. In this $\omega \rightarrow 0$ limit,
$\Sigma_K^B(\omega) \to \ \mbox{ct}
\in \Re$, and $\Sigma_R(\omega)$ is linear.
Therefore the noise kernels approach a classical Ohmic white-noise limit
with `temperature'
\begin{equation}
T^*=\lim_{\omega \rightarrow 0}\Sigma_K^B(\omega)
/[2\partial_{\omega}\Im\Sigma_R^B(\omega)]
\; . 
\end{equation}
At $V=0$ one gets
$T^*=T$. Instead, at $T=0$ and $eV \ll \hbar\omega_F$, one has
$T^*=eV/2$: the voltage plays the r{\^o}le of a {\it bath}
temperature. This fact has already been reported and it is at the root
of the derivation of the stochastic Gilbert equation for a spin under
bias \cite{Gilbert}. Having argued that the long-time dynamics
is governed by a {\it classical} Langevin equation at temperature $T^*$, it is justified that the
correlation scales as the one in~\cite{Cude} for $t/t_w=O(1)$, see
Eq.~(\ref{eq:C}), a result with two interesting consequences.  In the
case of (large $M$) quantum $3d$ coarsening the classical-quantum
mapping extends to space-time correlations~\cite{preparation} and
proves the existence of a growing coherence length $\xi(t_w)\propto
t_w^{1/2}$ over which the rotors are oriented in the same direction
and provides a real-space interpretation of aging.  Moreover, in the
same long-time regime, the linear response also scales as in the
classical problem. Therefore, the quantum fluctuation-dissipation
relation between integrated linear-response, $\chi$, and symmetric correlation
approaches the classical one, $\chi \sim \mbox{ct} + (q_{\rm
EA}-C)/T_{\rm eff}$, with an {\it infinite} effective
temperature~\cite{Cukupe}, $T_{\rm eff}\to\infty$, as shown in
Fig.~\ref{fig:C-chi}(b) (see also~\cite{Culo,other-quantum}). In
short, the asymptotic aging dynamics are universal, in the sense that
the scaling functions do not depend on $T,\Gamma,V$ in the coarsening
phase and, hence, are equivalent to the classical un-driven ones
($\Gamma=V=0$)~\cite{Cukulepe}.  This result mirrors the one obtained in
\cite{MitraKimMillis} for steady state dynamics.

The environment plays a dual r\^ole: its quantum character basically
determines the phase diagram but the coarsening process at long times
and large length-scales only `feels' a classical white bath at
temperature $T^*$. The two-time dependent decoherence phenomenon
(absence of oscillations, validity of a classical FDT when
$t/t_w=O(1)$, {\it etc.}) is intimately related to the development of
a non-zero (actually infinite) effective temperature, $T_{\rm eff}$,
of the system as defined from the deviation from the (quantum)
FDT~\cite{Cukupe}.  $T_{\rm eff}$ should be distinguished from $T^*$
as the former is generated not only by the environment but by the
system interactions as well ($T_{\rm eff}>0$ even at
$T^*=0$~\cite{Culo,other-quantum}). Moreover, we found an extension of
the irrelevance of $T$ in classical ferromagnetic coarsening ($T=0$
`fixed-point' scenario): after a suitable normalization of the
observables that takes into account all microscopic fluctuations ({\it
e.g.} $q_{\rm EA}$) the scaling functions are independent of all
parameters including $V$ and $\Gamma$.  Although we proved this result
through a mapping to a Langevin equation that applies to quadratic
models only, we expect it to hold in all instances with the same type
of ordered phase, say ferromagnetic, and a long-time aging dynamics
dominated by the slow motion of large domains. Thus, a large class of
coarsening systems (classical, quantum, pure and disordered) should be
characterized by the same scaling functions.  It could be worth
studying carefully systems evolving by barrier crossing, a rapid
process in which not only the low frequency behavior of the bath may
be relevant.

We thank C. Chamon, L. Chaput, A. Millis and A. Mitra for useful 
discussions. LFC is a member of IUF.


\begin{thebibliography}{99}
\bibitem{OnukiKawasaki} A. Onuki and K. Kawasaki, 
Ann. Phys. (N.Y.) {\bf 121}, 456 (1979).

\bibitem{DombGreen} B. Schmittmann and R. K. P. Zia, 
Vol. 17 of Phase Transitions and Critical  Phenomena,
eds. C. Domb and J. L. Lebowitz (Academic Press, London, 1995).

\bibitem{Struick} L. C. E. Struick, {\it Physical Aging 
in Amorphous Polymers and Other Materials}, (Elsevier, Amsterdam, 1978).

\bibitem{LeticiaLesHouches} L. F. Cugliandolo, 
in Les Houches Session 77, J-L. Barrat {\it et al}
eds. (Springer-EDP Sciences, 2002), arXiv:cond-mat/0210312. 
G. Biroli, J. Stat. Mech. P05014 (2005).

\bibitem{MitraMillisReichman} 
L. Arrachea and L. F. Cugliandolo, Europhys. Lett. {\bf 70}, 642 (2005). 
D. Segal {\it et al},
Phys. Rev. B {\bf 76}, 195316 (2007) and refs therein. 


\bibitem{DalidovichPhillips} D. Dalidovich and P. Phillips,
  Phys. Rev. Lett. {\bf 93}, 027004 (2004).

\bibitem{GreenSondhi} A. G. Green and S. L. Sondhi,
  Phys. Rev. Lett. {\bf 95}, 267001 (2005).

\bibitem{HoganGreen} P. M. Hogan and A. G. Green, arXiv:cond-mat/0607522.

\bibitem {MitraKimMillis}   A. Mitra {\it et al}, 
Phys. Rev. Lett. {\bf 97}, 236808 (2006).

\bibitem{Feldman}  D. E. Feldman, Phys. Rev. Lett. {\bf 95}, 177201 (2005).

\bibitem {MitraMillis} A. Mitra and A. J. Millis, arXiv:0804.3980

\bibitem{Ovadyahu} Z. Ovadyahu, Phys. Rev. B {\bf 73}, 214204 (2006).

\bibitem{Popovic} D. Popovic {\it et al.} Proceedings of SPIE, {\bf 5112}, 99 (2003).

\bibitem{mueller} E. Lebanon, M. M\"ueller, Phys. Rev. B {\bf 72}, 174202 (2005).

\bibitem{SachdevBook} S. Sachdev, {\it Quantum Phase Transitions}, 
(Cambridge Univ. Press, 1999).

\bibitem{SachdevYe} J. Ye, S. Sachdev, and N. Read, Phys. Rev. Lett. {\bf 70}, 4011 (1993).
T. K. Kope\'c, Phys. Rev. B {\bf 50}, 9963 (1994). 

\bibitem{Cude} L. F. Cugliandolo and D. S. Dean,
J. Phys. A {\bf 28}, 4213 (1995). 

\bibitem{Rokni}
M. Rokni and P. Chandra, Phys. Rev. B {\bf 69},  094403 (2004). 

\bibitem{Culo} L. F. Cugliandolo and G. S. Lozano, 
Phys. Rev. Lett. {\bf 80}, 4979 (1998); Phys. Rev. B {\bf 59}, 915 (1999). 

\bibitem{SchwingerKeldysh} U. Weiss, {\it Quantum dissipative systems}
(World Scientific, Singapore, 1993).  A. Kamenev, in 
Les Houches Session 81, H. Bouchiat {\it et al.} eds. 
(Springer-EDP Sciences, 2004), arXiv:cond-mat/0412296.

\bibitem{preparation} C. Aron, G. Biroli, and L. F. Cugliandolo, in 
preparation.

\bibitem{effect-bath}
L. F. Cugliandolo {\it et al}, 
Phys. Rev. B {\bf 66}, 014444 (2002).



\bibitem{Gilbert} A. S. N{\'u}{\~n}ez and R.
A. Duine, Phys. Rev. E {\bf 77}, 054401 (2008).

\bibitem{other-quantum} 
M. P. Kennett and C. Chamon, Phys. Rev. Lett. {\bf 86}, 1622 (2001).  
G. Biroli and O. Parcollet, Phys. Rev. B {\bf 65}, 094414 (2002).

\bibitem{Cukupe} L. F. Cugliandolo {\it et al}, 
Phys. Rev. E {\bf 55}, 3898 (1997). 

\bibitem{Cukulepe} Coarsening survives at finite $T$ 
under the current as opposed to the fact that 
aging is killed by a shear rate in the classical limit, see
L. F. Cugliandolo {\it et al}, Phys. Rev. Lett. {\bf 78}, 350 (1997)
and L. Berthier {\it et al} Phys. Rev. E {\bf 61}, 5464 (2000). We
attribute the difference to the fact that $T^*<+\infty$ in the quantum
model while the $T^*$ one can associate to shear
diverges~\cite{preparation}.

\end{thebibliography}
\end{document}